\documentstyle[preprint,aps]{revtex}
\begin{document}
\draft
\tightenlines
\preprint{INJE-TP-01-09, hep-th/0110234}
\title{ Check of the Mass Bound Conjecture in de Sitter Space }
\author{Rong-Gen Cai$^{1}$\footnote{email address: cairg@itp.ac.cn},
  Yun Soo Myung$^2$\footnote{email address: ysmyung@physics.inje.ac.kr}
and  Yuan-Zhong Zhang$^{3,1}$\footnote{email address: yzhang@itp.ac.cn}}

\address{ $^1$ Institute of Theoretical Physics, Chinese Academy of Sciences,
  \\
    P.O. Box 2735, Beijing 100080, China \\
 $^2$ Relativity Research Center and School of Computer Aided Science, \\
  Inje University, Kimhae 621-749, Korea \\
 $^3 $ CCAST (World Lab), P.O. Box 8730, Beijing 100080, China }

\maketitle
\begin{abstract}
Recently an interesting conjecture was put forward which states that any 
asymptotically de Sitter space whose mass exceeds that of exact de Sitter
space contains a cosmological singularity.  In order to test this mass bound 
conjecture,  we present two solutions.  One is the topological de Sitter solution 
and the other is its dilatonic deformation.  Although the latter is not 
asymptotically de Sitter space, the two solutions have a cosmological horizon and 
a cosmological singularity.  Using surface counterterm method we compute the 
quasilocal stress-energy tensor of gravitational field and the mass of the two 
solutions.  It turns out that this conjecture holds within  the two examples.
 Also we show that the thermodynamic quantities associated with the cosmological
horizon of the two solutions obey the first law of thermodynamics. Furthermore, the
nonconformal extension of dS/CFT correspondence is discussed.
\end{abstract}

\newpage

\section{Introduction}

It is  well-known  that to calculate the conserved charges including mass is
 a  difficult task in an asymptotically de Sitter (dS)
spacetime.  This is due to the absence of the spatial infinity and the globally timelike
Killing vector in  such a spacetime. In a recent paper \cite{BBM}, a novel prescription
was proposed for computing the boundary stress tensor and conserved charges of
asymptotically dS spacetimes from data at early or late time
infinity. This uses  the surface counterterm method \cite{BK,HS,KLS}, which
was developed in the AdS/CFT correspondence \cite{Mald,Gubs,Witt1}.
On the other hand,  if one accepts the dS/CFT correspondence \cite{stron1,stron2},
the resulting quantities then correspond to the stress-energy tensor and corresponding
conserved charges of the dual Euclidean conformal field theory (CFT).

Following  this prescription, the authors of \cite{BBM} calculated the
masses of the 3,4,5-dimensional
 Schwarzschild-de Sitter black hole solutions, respectively.
It is found that these masses are always less than those of  dS spaces in
corresponding dimensions. Furthermore,  they argued that this  result is
consistent with the dS/CFT correspondence and the Bousso's observation \cite{Bouss}
 on the asymptotically dS space  that the entropy of dS space 
is an upper bound for the entropy of any asymptotically dS space.
On the basis of this result, the authors of \cite{BBM} put
forward a conjecture (BBM conjecture):
 {\it Any asymptotically de Sitter space whose mass exceeds that of de Sitter
contains a cosmological singularity.}
Because a rigorous proof of this conjecture  is not yet carried out, it
is very interesting to check this conjecture with some examples. This is  the main 
aim of this paper.

The organization of this paper is as follows.
In Sec.~II we  introduce briefly the prescription to calculate
the boundary stress-energy tensor and  conserved charges of gravitational
field in the asymptotically de Sitter space. We present the
topological de Sitter solution and compute the  boundary stress-energy tensor and
 mass of this solution in Sec.~III. In Sec.~IV, we check the BBM conjecture in a  
 dilatonic deformation of the topological dS solution. We summarize  our results 
in Sec.~V with some discussions.


\section{ Prescription}
In this section we briefly review the surface counterterm method
to compute the conserved charges in asymptotically de Sitter space.
We consider an $(n+2)$-dimensional Einstein action with a positive cosmological
constant, $\Lambda =n(n+1)/2l^2$,
\begin{equation}
\label{eq1}
S=-\frac{1}{16\pi G}\int_{\cal M}d^{n+2}x\sqrt{-g}
   \left (R-\frac{n(n+1)}{l^2}\right) +\frac{1}{8\pi G}\int^{\partial \cal M^+}
  _{\partial \cal M^-}d^{n+1}x\sqrt{h}K.
\end{equation}
Here the first term is the bulk action with $n+2$-dimensional Newtonian constant
$G$. The second is the Gibbons-Hawking surface term, which is necessary to have
a well-defined Euler-Lagrange variation. ${\cal M}$ denotes the bulk manifold, 
$\partial {\cal M}^{\pm}$
are spatial boundaries at early and late times. $g_{\mu\nu}$ is the bulk metric and
$h_{ij}$ and $K$ are the induced metric and the trace of the extrinsic curvature
of the boundaries. In dS space the spacelike boundaries ${\cal I}^{\pm}$
are Euclidean surfaces at early and late time infinities. The notation $\int
^{\partial \cal M^+}_{\partial {\cal M}^-}d^{n+1}x$ indicates an integral over the
late time boundary minus an integral over the early time boundary which are both 
Euclidean surfaces.

Some surface counterterms have been given in \cite{BBM}, which can render the  action 
finite in 3,4,5-dimensional asymptotically dS spaces\footnote{The surface counterterms 
in the asymptotically dS space have also been discussed in \cite{Klem,Noji1,Noji2}.}
\begin{equation}
\label{eq2}
S_{\rm ct}= \frac{1}{8\pi G} \int_{\partial \cal M^+}d^{n+1}\sqrt{h}{\cal L}_{\rm ct}
    +\frac{1}{8\pi G}\int _{\partial \cal M^-}d^{n+1}x\sqrt{h}
   {\cal L}_{\rm ct},
\end{equation}
where
\begin{equation}
\label{eq3}
{\cal L}_{\rm ct}=\frac{n}{l}-\frac{l}{2(n-1)}{\cal R}
\end{equation}
and ${\cal R}$ is the intrinsic curvature of the induced metric.
This is an extension of the surface counterterm in  the asymptotically anti-de 
Sitter (AdS) space \cite{BK,HS,KLS}.
Decomposing the bulk spacetime in the ADM form as
\begin{equation}
\label{eq4}
ds^2 =g_{\mu\nu}dx^{\mu}dx^{\nu}=-N_t^2dt^2 +h_{ij}(dx^i +V^idt)(dx^j+V^jdt),
\end{equation}
one then has the induced metric $h_{ij}$ on spacelike surfaces of fixed time.
Denoting the future pointing unit normal to these surfaces by $u^i$, the extrinsic
curvature of these surfaces can be obtained using the formula:
\begin{equation}
K_{ij}=-h_i^{\ \mu}\nabla_{\mu}u_j.
\end{equation}
With these and the  Brown-York prescription \cite{BY}, one can get
the Euclidean quasilocal stress-energy tensor of an asymptotically dS space
\begin{eqnarray}
\label{eq6}
&& T^+_{ij}=\frac{2}{\sqrt{h}}\frac{\delta I}{\delta h^{ij}}
    =-\frac{1}{8\pi G}\left( K_{ij}-Kh_{ij}-\frac{n}{l}h_{ij} -\frac{l}{n-1}{\cal G}
     _{ij}\right), \nonumber \\
&& T^-_{ij}=\frac{2}{\sqrt{h}}\frac{\delta I}{\delta h^{ij}}
    =-\frac{1}{8\pi G}\left( -K_{ij}+Kh_{ij}-\frac{n}{l}h_{ij} -\frac{l}{n-1}{\cal G}
     _{ij}\right).
\end{eqnarray}
Here $I=S +S_{\rm ct}$, and ${\cal G}_{ij}$ is the Einstein tensor of the induced
surface. Since there exist two spacelike boundaries in dS space, the superscripts
$\pm$ in  $T_{ij}$ represent the quantity on
the late or early time boundary. The difference in signs of the two stress-energy tenors
in (\ref{eq6}) arises because the extrinsic curvature $K$ is defined with respect
to a future pointing timelike normal, leading to sign changes between the early
and late time boundaries~\cite{BBM}. For this reason as in \cite{BBM} we will 
 use $T_{ij}=T_{ij}^+$ in what follows.  This means that we  calculate the conserved 
charges on the late time boundary ${\cal I}^+$.

Next let us decompose the induced metric $h_{ij}$ in the form
\begin{equation}
\label{eq7}
h_{ij}dx^idx^j = N^2_{\rho}d\rho^2 +\sigma_{ab}(d\phi^a +N^a_{\Sigma}d\rho)
    (d\phi^b +N^b_{\Sigma}d\rho),
\end{equation}
where the notation $\phi^a$ are angular variable parameterizing closed surfaces
around the origin. Suppose $\xi^i$ to be a Killing vector generating an isometry of the
boundary geometry. Following \cite{BY,BK}, one can define the conserved charge $Q$
associated with the Killing vector $\xi^i$ using the quasilocal stress-energy
tensor $T_{ij}$ as follows
\begin{equation}
\label{eq8}
Q=\oint_{\Sigma}d^n\phi \sqrt{\sigma}n^i\xi^jT_{ij},
\end{equation}
where $n^i$ is the unit normal to the surface $\Sigma$ with a fixed $\rho$, and the coordinate
$\rho$ is obtained by analytic continuation of a timelike Killing vector.

Recall that an important obstacle to define the mass of gravitational field in the
asymptotically dS space is the absence of a globally timelike Killing vector. However,
 there is a Killing
vector which is timelike within the cosmological horizon of dS space in the static
coordinates, while it is spacelike outside the cosmological horizon and then on
 ${\cal I}^{+}$, future null infinity. Thus any spacetime which is
asymptotically dS space will have such an asymptotic symmetry generator. Adapting
the coordinates (\ref{eq7}) so that ``radial'' normal $n^i$ is proportional to the
relevant (spacelike) boundary Killing vector $\xi^i$, the authors of \cite{BBM} 
proposed a mass formula for asymptotically dS spaces:  
\begin{equation}
\label{eq9}
M=\oint_{\Sigma}d^n\phi \sqrt{\sigma}N_{\rho}\epsilon, \ \ \
    \epsilon \equiv n^in^jT_{ij}.
\end{equation}
Here the Killing vector $\xi^i$ is normalized as $\xi^i=N_{\rho}n^i$. Similarly the
angular momenta can be defined as
\begin{equation}
\label{eq10}
P_a=\oint_{\Sigma}d^n\phi \sqrt{\sigma}{\cal J}_a, \ \ \
   {\cal J }_a=\sigma_{ab}n_iT^{bi}.
\end{equation}

Using this prescription,
 the masses of 3, 4, 5-dimensional Schwarzschild-dS black hole solutions have been calculated in
\cite{BBM}. It was found that the mass of  dS space is always larger than
that of the black hole solution in the dS space in corresponding
dimensions. This leads to
the BBM conjecture. Now we wish to check this conjecture with the
following two solutions.


\section{Topological de Sitter solution}

We start with an ($n+2$)-dimensional topological black hole solution in AdS space
\begin{equation}
\label{eq11}
ds^2_{TBAdS}=-f(r)dt^2 +f(r)^{-1}dr^2 +r^2\tilde g_{ab}dx^adx^b,
\end{equation}
where
\begin{equation}
f(r) = k-\frac{2Gm}{r^{n-1}}+\frac{r^2}{l^2},\ \ \ k=1,-1,0.
\end{equation}
$\tilde g_{ab} dx^adx^b$ is the line element of an $n$-dimensional hypersurface with 
constant curvature $kn(n-1)$  and volume $V =\int d^nx \sqrt{\tilde g}$. $l$
is the curvature radius of AdS space.  $m$ is a constant related with the ADM mass of 
the black hole\cite{BIRM}.  It is believed that  black holes in asymptotically flat
spacetime should have a spherical horizon. When there is a  
negative cosmological constant in a spacetime,  however, a black hole can have
a non-spherical horizon. In this sense  this  black hole (\ref{eq11}) is referred to as
a topological black hole in AdS space. When $m=0$, the solution (\ref{eq11}) reduces to 
the  AdS space. Replacing $l^2$ by $-l^2$ in (\ref{eq11}), one has a solution 
\begin{equation}
\label{eq13}
ds^2_{TBdS}=-f(r)dt^2 +f(r)^{-1}dr^2 +r^2\tilde g_{ab}dx^adx^b,
\end{equation}
where
\begin{equation}
\label{eq14}
f(r) = k-\frac{2Gm}{r^{n-1}}-\frac{r^2}{l^2}, \ \ \ k=1,-1,0.  
\end{equation}
Obviously this is a solution to the Einstein equations with a positive cosmological
constant in ($n+2$) dimensions. 

When $k=1$, it is just the Schwarzschild-de Sitter solution. The case $m=0$ reduces
to the dS space with a cosmological horizon $r_c=l$. 
 When $m$ increases, a black
hole horizon occurs and increases with the size of $m$, while the cosmological
horizon shrinks. Finally the black hole horizon touches the cosmological horizon
when
\begin{equation}
\label{eq15}
m =\frac{1}{G(n+1)}\left(\frac{n-1}{n+1}l^2\right)^{(n-1)/2}.
\end{equation}
This is  the Nariai black hole,  the maximal black hole in dS space. The mass of the 
solution in this case has been calculated in \cite{BBM}.  We will discuss the
cases $k=0$ and $k=-1$, respectively.

 {\it (i) The case of $k=0$}. In this case, $\tilde g_{ab}dx^adx^b$ is an
$n$-dimensional Ricci flat hypersurface.  Changing the sign in front of $m$ in 
(\ref{eq14}), one has 
\begin{equation}
\label{eq16}
ds^2_{TdS} =-\left(\frac{2Gm}{r^{n-1}} -\frac{r^2}{l^2}\right)dt^2 +
 \left(\frac{2Gm}{r^{n-1}} -\frac{r^2}{l^2} \right)^{-1}dr^2 + r^2dx_n^2, 
\end{equation}
where $dx_n^2$ denotes the Ricci flat hypersurface.
It is easy to check that the metric (\ref{eq16}) is still a solution to the Einstein
equations with a positive cosmological constant in ($n+2)$ dimensions.
 From this solution
we  see that there is a Ricci flat cosmological horizon at $r=r_c=(2Gml^2)^{1/(n+1)}$.
Also there exists a cosmological singularity at $r=0$ for $n \ge 2$ and $m\ne 0$.
Therefore, this  solution is a good example to check the BBM conjecture.
For this cosmological
horizon, we  have the Hawking temperature $T_{\rm HK}$ and entropy $S$,
\begin{eqnarray}
\label{eq17}
&& T_{\rm HK}= \frac{(n+1)r_c}{4\pi l^2}, \nonumber \\
&& S =\frac{r_c^n V}{4G}.
\end{eqnarray}
When $m=0$, the solution (\ref{eq16}) goes to
\begin{equation}
\label{eq18}
ds^2 = -\frac{l^2}{r^2}dr^2 +\frac{r^2}{l^2} dt^2 +r^2 dx_n^2,
\end{equation}
in which $t(r)$ becomes a spacelike (timelike) coordinate. In fact, this is a
pure dS space: One can rewrite the metric (\ref{eq18}) as follows,
\begin{equation}
\label{eq19}
ds^2 = -d\tau^2 +e^{\pm 2\tau/l}dx_{n+1}^2,
\end{equation}
where $ dx_{n+1}^2$ is an $(n+1)$-dimensional Ricci-flat space. This is just the dS
space in the planar coordinates.

We now calculate the boundary stress-energy tensor and the mass of the solution
(\ref{eq16}). For $r>r_c$, this solution
can be rewritten as
\begin{equation}
\label{eq20}
ds^2 =-f(r)^{-1}dr^2 +f(r)dt^2 + r^2 dx_n^2, \ \  f=\frac{r^2}{l^2}
-\frac{2Gm}{r^{n-1}} >0
\end{equation}
in which $t(r)$ becomes a spacelike (timelike) coordinate.
Since $dx_n^2$ is a Ricci-flat space, therefore, for a hypersurface with a fixed $r>r_c$,
its induced metric, $f(r)dt^2 + r^2 dx_n^2$, is also Ricci flat. Thus those counterterms
involving the intrinsic curvature and Ricci tensor of the induced metric vanish, and 
the Lagrangian of the required surface counterterm for the solution (\ref{eq16}) is 
\begin{equation}
\label{eq21}
{\cal L}_{\rm ct}= \frac{n}{l},
\end{equation}
and the boundary stress-energy tensor becomes
\begin{equation}
\label{eq22}
T_{ij}=-\frac{1}{8\pi G} \left(K_{ij}-Kh_{ij}-\frac{n}{l}h_{ij}\right).
\end{equation}
Considering a surface $\Sigma$ with fixed $r>r_c$ in (\ref{eq20})  and calculating its
extrinsic curvature $K_{ij}$,  we obtain from (\ref{eq22})
\begin{eqnarray}
\label{eq23}
&& T_{tt}=\frac{nm}{8\pi lr^{n-1}} +\cdots, \nonumber\\
&& T_{ab}=-\frac{ml}{8\pi r^{n-1}}\delta_{ab} +\cdots,
\end{eqnarray}
where the ellipses denote higher order terms, which have no contribution when we take the
limit $r\to \infty$ on the  ${\cal I}^{+}$.

 Now we are in  a position to calculate the mass of the solution (\ref{eq16}).
Substituting this boundary stress-energy
tensor (\ref{eq23}) into the mass formula (\ref{eq9}), we find
\begin{equation}
\label{eq24}
M = \frac{nmV}{8\pi}.
\end{equation}
When $n=1$ and identifying the coordinate $x$ in (\ref{eq16}) with a circle with
period $2\pi$, from (\ref{eq24}) one has $M=m/4$, precisely reproducing the result
in Ref.~\cite{BBM} for the mass of three dimensional Schwarzschild-dS 
solution\footnote{Note
the difference of notations used in this paper and in Ref.~\cite{BBM}: $m_{\rm here}
=4m_{\rm there}$.}.
 When $m=0$, we have $M=0$. This is consistent with the result obtained 
in \cite{BBM,stron1} that
the mass of the three-dimensional dS space vanishes in the planar coordinates
\footnote{In the static coordinates, however, the three-dimensional dS space has 
a nonvanishing mass $M=1/8G$\cite{stron2,myung}.}. Our
result (\ref{eq24}) indicates
that the mass  of  dS space vanishes ($M_{\rm dS}=0$) in
any dimension in the planar coordinates. When $m \ne 0$, we have $M > M_{\rm dS}=0$.
According to the BBM conjecture\cite{BBM}, there should be a cosmological singularity.
Indeed it is clear from (\ref{eq16}) that there is a cosmological singularity at $r=0$.
As a result, we verify that the BBM conjecture holds in the  solution (\ref{eq16}).  
Furthermore, we can easily check that  the mass $M$ in (\ref{eq24}),
 Hawking temperature $T_{\rm HK}$ and entropy $S$ in (\ref{eq17}) satisfy the first law
of thermodynamics
\begin{equation}
\label{eq25}
dM = T_{\rm HK} dS.
\end{equation}

Next we calculate the stress-energy tensor of the Euclidean CFT dual to the solution
(\ref{eq16}). As the asymptotically  AdS case, the induced metric diverges when the 
boundary of dS space is 
approached\footnote{In the asymptotically AdS case, 
the induced metric also diverges when the boundary of AdS space is
approached. But a well-defined surface metric on which the dual CFT resides can be
determined, up to a conformal factor, from the bulk metric. The behavior
of induced metric near the boundary, the normalizations of action and boundary 
stress-energy tensor, and the gravitational conformal anomaly have been analyzed 
in detail in \cite{Haro}. For the asymptotically dS case, a similar analysis can be 
made as well, for example, see \cite{Mazu}.}. However,
 a surface metric on which the Euclidean CFT resides can be
fixed, up to a conformal factor, from the bulk metric (\ref{eq20}). For example,
a simple surface metric can be obtained as follows,  
\begin{equation}
\label{eq26}
ds^2_{ECFT}=\gamma_{ij}dx^idx^j = \lim_{r\to \infty}
\frac{l^2}{r^2}ds^2_{TdS}
     = dt^2 +l^2 dx_n^2.
\end{equation}
Note that here $t$ is a spacelike coordinate.
The stress-energy tensor $\tau_{ij}$
 of the boundary Euclidean CFT can be obtained using the following 
relation \cite{Myers}
\begin{equation}
\label{eq27}
\sqrt{\gamma}\gamma^{ij}\tau_{jk}=\lim_{r \to \infty} \sqrt{h}h^{ij}T_{jk}.
\end{equation}
Substituting (\ref{eq23}) into the above and using (\ref{eq26}), one has
\begin{eqnarray}
&& \tau_{tt}=\frac{nm}{8\pi l^n},    \nonumber \\
&& \tau_{ab}=-\frac{m}{8\pi l^{n-2}}\delta_{ab}.
\end{eqnarray}
As  expected, the trace of the stress-energy tensor vanishes\footnote{In general,
there is a conformal anomaly for a CFT in even 
dimensions (for a review see \cite{Duff}).  In the case we are 
discussing, however, the spacetime background (\ref{eq26}) is Ricci flat, 
therefore the conformal anomaly vanishes.}. 

{\it (ii) The case of $k=-1$.} In this case, changing the sign in front of $m$
in (\ref{eq13}), we have 
\begin{equation}
\label{Ieq1}
ds^2_{TdS} =-f(r)dt^2 +f(r)^{-1}dr^2 +r^2 \tilde g_{ab}dx^adx^b,
\end{equation}
where
\begin{equation}
\label{Ieq2}
f(r)=-1 +\frac{2Gm}{r^{n-1}} -\frac{r^2}{l^2}.
\end{equation}
Once again, when $m>0$, this solution has a cosmological singularity at $r=0$ and 
a cosmological horizon $r_c$ which is a negative constant curvature hypersurface. 
In this sense
we refer to the solution (\ref{Ieq1}) together with the solution (\ref{eq16}) as
 topological dS solutions.  The cosmological horizon of the solution (\ref{Ieq1})
has the Hawking temperature $T_{\rm HK}$ and entropy $S$
\begin{eqnarray}
\label{Ieq3}
&& T_{\rm HK}= \frac{1}{4\pi r_c}\left (n-1 +(n+1)\frac{r_c^2}{l^2}\right),
   \nonumber \\
&& S=\frac{r_c^n V}{4G}.
\end{eqnarray}
Since the $\tilde g_{ab}dx^adx^b$ is a negative constant curvature hypersurface, in this
case, so the surface counterterm (\ref{eq21}) is not sufficient.
The needed surface counterterms will depend on the spacetime dimension. 
We consider therefore the four- and five- dimensional cases below. In that case, the
required surface counterterms are given in (\ref{eq3}), and corresponding boundary
stress-energy tensor can be computed using (\ref{eq6}).  

Repeating the steps as the case of $k=0$ and using (\ref{eq6}), in four dimensions, 
we obtain the boundary stress-energy tensor
\begin{eqnarray}
\label{Ieq4}
&& T_{tt}= \frac{m}{4\pi rl} +\cdots, \nonumber \\
&& T_{ab}=-\tilde{g}_{ab} \frac{ml}{8\pi r}+\cdots. 
\end{eqnarray}
The mass of the solution is
\begin{equation}
\label{Ieq5}
M_4 =\frac{mV}{4\pi}.
\end{equation}
And the stress-energy tensor of corresponding Euclidean CFT is
\begin{eqnarray}
\label{Ieq6} 
&& \tau_{tt}=\frac{m}{4\pi l^2}, \nonumber \\
&&\tau_{ab}=-\tilde g_{ab}\frac{m}{8\pi},
\end{eqnarray}
which has a vanishing trace. The CFT resides on the three dimensional space 
with metric
\begin{equation}
\label{Ieq7}
\gamma_{ij}dx^idx^j = dt^2 +l^2 \tilde{g}_{ab}dx^adx^b.
\end{equation}

In five dimensions, we find that it is quite different from the case of four 
dimensions. The boundary stress-energy tensor is
\begin{eqnarray}
\label{Ieq8}
&& T_{tt}= \frac{3l}{8\pi G r^2}\left(\frac{1}{8} +\frac{Gm}{l^2}
    \right) +\cdots, \nonumber \\
&& T_{ab}=-\tilde{g}_{ab}\frac{l^3}{8\pi G r^2}\left(\frac{1}{8}
     +\frac{Gm}{l^2}\right) +\cdots. 
\end{eqnarray}
The mass of the solution is
\begin{equation}
\label{Ieq9}
M_5 =\frac{3V}{8\pi G}\left(\frac{l^2}{8} +Gm\right),
\end{equation}
and the stress-energy tensor of Euclidean CFT 
\begin{eqnarray}
\label{Ieq10}
&& \tau_{tt}=\frac{3}{8\pi Gl}\left(\frac{1}{8}+\frac{Gm}{l^2}\right) , 
    \nonumber \\
&& \tau_{ab}=-\tilde g_{ab}\frac{l}{8\pi G}\left(\frac{1}{8}+\frac{Gm}{l^2}
    \right).
\end{eqnarray}
The surface metric is of the form (\ref{Ieq7}), but in four dimensions. Once again, 
the stress-energy tensor has a vanishing trace\footnote{In four dimensions, in general
there is a gravitational conformal anomaly \cite{Duff,HS,Noji1} proportional 
to $({\cal R}_{ij}{\cal R}^{ij}-
{\cal R}^2/3)$, where ${\cal R}$ and ${\cal R}_{ij}$ are curvature scalar and Ricci
tensor of surface metric, respectively.
 For the spacetime background (\ref{Ieq7}), however, it is easy check that
this conformal anomaly vanishes. This explains why the stress-energy tensor of dual CFT
has a vanishing trace.}.    From
(\ref{Ieq9}) we see that  unlike the case (\ref{Ieq5}) in four dimensions,
the mass in five dimensions does not vanish even when $m=0$. This is reminiscent 
of the difference between the four and five dimensional Schwarzschild-dS black hole 
solutions \cite{BBM}, there it is also found that there is a nonvanishing mass
for the five dimensional pure dS space.  But from (\ref{Ieq7}) and (\ref{Ieq9}) we 
see that the masses with
$m>0$  are always larger than those of dS solutions with $m=0$. As a result,
 we confirm the BBM conjecture in the topological dS solution. Furthermore, 
these masses (\ref{Ieq7}) and (\ref{Ieq9}) also satisfy the first law (\ref{eq25}) of
thermodynamics.


\section{Dilatonic solution}

The second example  is a dilatonic deformation of the topological dS solution with
a Ricci flat cosmological horizon (\ref{eq16}).
Consider the following action of a dilaton gravity theory,
\begin{equation}
\label{eq29}
S =-\frac{1}{16\pi G}\int_{\cal M}d^{n+2}x\sqrt{-g}\left(R -\frac{1}{2}
(\partial \phi)^2
   +V_0e^{-a \phi}\right),
\end{equation}
where $a$ and $V_0$ are  assumed to be two positive constants.
This action is an effective one for some gauged supergravity theories
\cite{BST,CO}.
In \cite{CZ} (see also \cite{CZ1,CJS} for the case in
four dimensions), a class of domain wall black hole solutions has been found,
\begin{eqnarray}
\label{eq30}
&& ds^2_{DB}= -f(r)dt^2 +f(r)^{-1}dr^2 +R^2dx_n^2, \nonumber \\
&& R(r)=r^N, \ \ \phi(r)=\phi_0 +\sqrt{2nN(1-N)} \ln r, \nonumber \\
&& f(r)= \frac{V_0 e^{-a\phi_0}r^{2N}}{nN(N(n+2)-1)}-\frac{m r^{1-nN}}
           {\sqrt{2nN(1-N)}},
\end{eqnarray}
where $\phi_0$ and $m$ are two integration constants.
Also the notation $dx_n^2$
denotes the line element of an $n$-dimensional Ricci-flat space
 and the parameter $N$ obeys the relation
\begin{equation}
\label{eq31}
a=\frac{\sqrt{2nN(1-N)}}{nN}.
\end{equation}
From (\ref{eq31}) and (\ref{eq30}) one can see that $N$ must satisfy $1/(n+2) <N\le 1$.
For $N=1$, the solution (\ref{eq30}) precisely recovers the topological black hole 
with a Ricci flat horizon in AdS space\cite{BIRM}.
For a general $N$, the solution (\ref{eq30}) is  neither
asymptotically AdS, nor asymptotically flat. When $m=0$, the solution describes
a domain wall spacetime where a domain wall/QFT (quantum field theory)
correspondence \cite{BST}, including the AdS/CFT correspondence in the horospherical 
coordinates as a special case, arises: 
 a certain gauged supergravity on the domain wall spacetime is dual to
a QFT residing on the domain wall. For details, see \cite{BST,CO,CZ}.
It has been shown in \cite{CO} that  one can also get a well-defined
boundary stress-energy tensor by adding an appropriate surface counterterm for
a class of solutions like (\ref{eq30}), even though those solutions are not
asymptotically AdS. The quasilocal  boundary stress-energy tensor of the
gravitational field and therefore the stress-energy tensor of corresponding QFT
for the solution (\ref{eq30}) have been acquired in \cite{CZ}. It turns out that
the integration constant  $m$ in (\ref{eq30}) is proportional to the mass $M$ of 
the black hole\cite{CZ}:
\begin{equation}
\label{eq32}
M=\frac{nN}{\sqrt{2nN(1-N)}}\frac{mV}{16\pi G},
\end{equation}
where $V$ is the volume of the Ricci flat space $dx_n^2$ in (\ref{eq30}).

We now turn to the case $V_0<0$ in the action (\ref{eq29}). In this case, the action
still can  be viewed as an effective truncation of a certain gauged supergravity,
for example, see \cite{Town}.  For  convenience, we make a replacement: $V_0 \to
-V_0$ in (\ref{eq29}) so that we still have $V_0>0$ in the following. Then it
is easy to check the new action has still a solution like (\ref{eq30}), but
with a new function $f$:
\begin{equation}
\label{eq33}
f(r)= \frac{mr^{1-nN}}{\sqrt{2nN(1-N)}} -\frac{V_0e^{-a\phi_0}r^{2N}}
   {nN(N(n+2)-1)},
\end{equation}
and others keep unchanged. Here we have changed the sign in front of the
integration constant $m$ in (\ref{eq30}). For the case of $N=1$,
the new solution (\ref{eq33}) reduces to the
 $k=0$ topological dS solution considered above. For a general $N<1$, this 
solution is not asymptotically de Sitter. But it has a cosmological horizon $r_c$
\begin{equation}
\label{eq34}
r_c =\left(\frac{nNm(N(n+2)-1)}{V_0e^{-a\phi_0}\sqrt{2nN(1-N)}}
   \right)^{1/((n+2)N-1)}.
\end{equation}
And thus  it has associated Hawking temperature $T_{\rm HK}$ and entropy $S$ as
\begin{eqnarray}
\label{eq35}
&& T_{\rm HK}= \frac{V_0e^{-a\phi_0}r_c^{2N-1}}{4\pi nN}, \nonumber\\
&& S=\frac{ r_c^{nN}V}{4G}.
\end{eqnarray}
Furthermore, we note that there exists a cosmological singularity at $r=0$ in the
solution (\ref{eq33}). 
Although the solution is not asymptotically de Sitter,
 we find that one can get a well-defined
 quasilocal stress-energy tensor of gravitational field for
the solution (\ref{eq33}) by adding an appropriate surface counterterm to the bulk
action. The surface counterterm is given by
\begin{equation}
\label{eq36}
{\cal L}_{\rm ct}=\frac{n}{l_{\rm eff}} \sqrt{\frac{N(n+1)}{N(n+2)-1}},
 \ \ \  \frac{1}{l_{\rm eff}}=\sqrt{\frac{V_0e^{-a\phi}}{n(n+1)}}.
\end{equation}
And then the boundary stress-energy tensor is
\begin{equation}
\label{eq37}
T_{ij}=-\frac{1}{8\pi G} \left (K_{ij} -Kh_{ij}-
    \frac{n}{l_{\rm eff}} \sqrt{\frac{N(n+1)}{N(n+2)-1}}h_{ij} \right).
\end{equation}
Similarly we
obtain the boundary stress-energy tensor of gravitational field on the 
surface $\Sigma$ with a fixed $r>r_c$,
\begin{eqnarray}
\label{eq38}
&& T_{tt}=\frac{nNm c^{1/2} r^{-(n-1)N}}{16\pi G \sqrt{2nN(1-N)}} +\cdots,
   \nonumber \\
&& T_{ab} = -\delta_{ab} \frac{(2N-1)m r^{-(n-1)N}}{16\pi G c^{1/2}\sqrt{2nN(1-N)}}
   +\cdots,
\end{eqnarray}
where the ellipses represent higher order terms which have no contribution when we take 
the limit $r\to \infty$  on ${\cal
I}^{+}$. The constant $c$ is
\begin{equation}
\label{eq39}
c = \frac{V_0e^{-a\phi_0}}{nN(N(n+2)-1)}.
\end{equation}
Using the mass formula (\ref{eq9}) and the boundary stress-energy tensor (\ref{eq38}),
we obtain the mass of the solution (\ref{eq33})
\begin{equation}
\label{eq40}
M = \frac{nNmV}{16\pi G\sqrt{2nN(1-N)}}.
\end{equation}
We find that the mass (\ref{eq40}) of the dilatonic deformation (\ref{eq33})
 has the same form as the domain-wall black hole solution (\ref{eq30}).  
Furthermore, we can see that the mass
in (\ref{eq40}), the Hawking temperature $T_{\rm HK}$ and entropy in (\ref{eq35})
 satisfy the first law (\ref{eq23}) of thermodynamics. From (\ref{eq40}) one has
  $M_{\rm vac}=0$, for the vacuum state ($m=0$) in the solution
(\ref{eq33}). Thus $M >M_{\rm vac}=0$, showing that the BBM conjecture is also 
satisfied with  the dilatonic
deformation of the topological de Sitter solution, even though  the solution is
not asymptotically de Sitter.

Since the solution (\ref{eq33}) is not asymptotically de Sitter, we do not expect
that the dual is a Euclidean CFT. Instead we expect that there is  a  Euclidean QFT
dual to the solution (\ref{eq33}). This
correspondence is an analog of the domain wall/QFT correspondence in the spacetime
with a cosmological horizon. In the correspondence,
 we  can calculate the stress-energy tensor of the QFT dual to the 
 solution (\ref{eq33}). As the asymptotically  dS case, the surface
 metric $\gamma_{ij}$ of the spacetime, on which
the QFT resides, can be determined, up to a conformal factor, as follows,
\begin{equation}
\label{eq41}
ds^2_{EQFT}=\gamma_{ij}dx^idx^j =\lim_{r \to \infty} \frac{1}{r^{2N}}ds^2
   = cdt^2 +dx_n^2,
\end{equation}
where $t$ is a spacelike coordinate and $c$ is given in (\ref{eq39}). Using
(\ref{eq38}),
we obtain
\begin{eqnarray}
&& \tau_{tt}=\frac{nNmc^{1/2}}{16\pi G\sqrt{2nN(1-N)}},    \nonumber \\
&& \tau_{ab}=-\delta_{ab}\frac{(2N-1)m}{16\pi Gc^{1/2}\sqrt{2nN(1-N)}}.
\end{eqnarray}
As expected, its trace does not vanish unless $N=1$. In the case of $N=1$,
 the solution (\ref{eq33}) reduces to the $k=0$  topological dS solution, to which
 one has a Euclidean CFT dual. Note that for the $N=1$ case, those ill-defined
expressions can be remedied by redefining the integration constant $m$: 
for example, one can absorb the factor $\sqrt{1-N}$ into the $m$.


\section{Conclusion and Discussion}

The dS space  is the unique maximally symmetric curved spacetime. It enjoys the same 
degree of symmetry as Minkowski space. So it has been most
studied by quantum field theorists (see \cite{BD} and references therein). On the
other hand, the recent astronomical data of supernova \cite{Per,CDS,Gar} together
with the need of the inflation model in the cosmology of early universe indicate
that our universe approaches dS geometries in both the far past and the far 
future \cite{HKS,FKMP,Strom3}. Moreover, it has been proposed recently 
that there is a dual between quantum gravity on a dS space and a Euclidean
CFT residing on a boundary of the dS space \cite{stron1,stron2}, very like
the AdS/CFT correspondence.
Therefore  both the dS space itself and the realistic 
universe motivate us to well understand the dS space (including asymptotically
dS spaces). As a first step, one has to compute some conserved charges like
mass and angular momentum associated with asymptotically dS spaces. However, as
stated in INTRODUCTION, it is not an easy matter to obtain those conserved 
charges in asymptotically dS spaces because of the absence of spatial infinity
and globally timelike Killing vector. In Ref.~\cite{BBM} a novel prescription
has been proposed to calculate those conserved charges from data at early
or late time infinity. And it has been found that the masses of pure dS spaces
are always larger than those of Schwarzschild-dS black holes in the corresponding
dimensions. The interesting result is consistent with the 
observation \cite{Bouss} that 
the entropy of pure dS space is an upper bound of any asymptotically dS spaces if
one accepts the dS/CFT correspondence, because generically field theories should have 
entropies increasing with energy. The interesting result also leads to the 
BBM conjecture.  If the BBM conjecture is correct, no doubt it is a very important
feature of asymptotically dS spaces. 

In order to check the BBM conjecture, we have presented two solutions, the topological 
dS solution and
its dilatonic deformation. Both have a cosmological horizon and a cosmological
singularity. Using surface counterterm method, we have calculated the boundary quasilocal
stress-energy tensor of gravitational field and obtained the masses for both solutions.
The resulting masses are always positive, while the masses of dS space and its dilatonic
deformation vanish in the planar coordinates. Although the mass (\ref{Ieq9}) of dS 
space in five dimensions does not vanish for the case $k=-1$, in all cases we considered
in this paper, we  have verified  the BBM conjecture. Thus we have provided
evidence in favor of the BBM conjecture.

Even though the dilatonic deformation of the topological dS solution is not
asymptotically de Sitter, we expect that there exists a dual
 Euclidean QFT. This correspondence is considered as an analog of the
domain wall/QFT correspondence \cite{BST}: quantum gravity on the background
(\ref{eq33}) is dual to a certain Euclidean QFT residing on the space (\ref{eq41}).
Thus we can view this correspondence as a Euclidean version of  the domain wall/CFT
correspondence.
The Euclidean domain wall/QFT correspondence includes the dS/CFT correspondence 
in the planar coordinates as a special case, in the same way as that the
AdS/CFT correspondence in the horospherical coordinates comes out as a special case 
in the domain wall/QFT correspondence \cite{BST}. This Euclidean domain wall/QFT 
correspondence is a nonconformal extension of the dS/CFT correspondence. 
According to this correspondence, 
we have obtained the stress-energy tensor of corresponding Euclidean QFTs. Also we 
have calculated some thermodynamic quantities associated with the cosmological horizon 
of these solutions, and verified that they all obey the first law of thermodynamics.

If the BBM conjecture indeed holds, then an interesting question is what its implications
are. The BBM conjecture says that {\it any asymptotically de Sitter space whose 
mass exceeds that of pure de Sitter space contains a cosmological singularity.}
From another point of view to see it, it gives us an upper bound of mass
for any asymptotically dS space without a cosmological singularity. Together with 
the dS/CFT correspondence and the Bousso's observation about the maximal entropy bound
for any asymptotically dS space, the BBM conjecture seems to imply that the maximal 
entropy bound must be violated in any asymptotically dS space with a cosmological 
singularity. Thus, the BBM conjecture further implies that some energy conditions must
be also violated for asymptotically dS space with a cosmological singularity. Therefore
the BBM conjecture might be quite useful to well understand the creation and fate of our
realistic universe. In addition, there might exist CFTs dual to some asymptotically dS spaces
with cosmological singularity. For such CFTs, there is no maximal entropy bound. Some
energy conditions might be violated for these exotic CFTs. Therefore a well understanding of
the BBM conjecture will provide help to establish the dS/CFT correspondence. However, 
the understanding so far gained to the conjecture is obviously incomplete.  It is to be 
expected to reveal the deep implications of the conjecture in a near future.

\section*{Acknowledgments}
RGC thanks M. Yu for a useful discussion.
The work of RGC  was supported in part by a grant from Chinese Academy of
Sciences. YSM was supported
in part by the Brain Korea 21 Program, Ministry of Education, Project No.
D-0025 and KOSEF, Project No. 2000-1-11200-001-3.
 YZZ was supported in part by National Natural
Science Foundation of China under Grant No. 10047004, and by Ministry of
Science and Technology of Modern Grant No. NKBRSF G19990754.


\end{document}